\documentclass[pdflatex,sn-mathphys-num]{sn-jnl}

\usepackage{graphicx}%
\usepackage{multirow}%
\usepackage{amsmath,amssymb,amsfonts}%
\usepackage{amsthm}%
\usepackage{mathrsfs}%
\usepackage[title]{appendix}%
\usepackage{xcolor}%
\usepackage{textcomp}%
\usepackage{manyfoot}%
\usepackage{booktabs}%
\usepackage{algorithm}%
\usepackage{algorithmicx}%
\usepackage{algpseudocode}%
\usepackage{listings}%

\theoremstyle{thmstyleone}%
\theoremstyle{thmstyletwo}%

\theoremstyle{thmstylethree}%

\begin{document}

\title[Noise-Driven Instrument]{Noise-Driven Instrument Based on Coherent Quantum and Stochastic Oscillator Models}

\author*[1]{\fnm{Felipe} \sur{González de la Maza}}\email{felipe.gonzalez@alumni.icfo.eu}

\author[1,4]{\fnm{Maciej} \sur{Lewenstein}}

\author[3]{\fnm{Antoine} \sur{Reserbat-Plantey}}

\author[1,2]{\fnm{Reiko} \sur{Yamada}}

\affil[1]{\orgname{ICFO -- Institut de Ciències Fotòniques, The Barcelona Institute of Science and Technology},
\orgaddress{\street{Av. Carl Friedrich Gauss 3}, \postcode{08860}, \city{Castelldefels (Barcelona)}, \country{Spain}}}

\affil[2]{\orgname{ESMUC – Escola Superior de Música de Catalunya},
\orgaddress{\city{Barcelona}, \country{Spain}}}

\affil[3]{\orgname{Université Côte d'Azur, CNRS, CRHEA},
\orgaddress{\city{Valbonne}, \country{France}}}

\affil[4]{\orgname{ICREA},
\orgaddress{\street{Pg. Lluís Companys 23}, \postcode{08010}, \city{Barcelona}, \country{Spain}}}

\abstract{In recent years, emerging research at the intersection of quantum physics and sound synthesis has opened new conceptual and technical possibilities for instrument design and sonic exploration. This study investigates the potential of formal analogies between quantum systems and classically non-deterministic systems for the generation of tangible acoustic phenomena. Specifically, it explores how quantum mechanical concepts can serve not only as metaphors but as operative frameworks in the design of new musical tools. Building on recent theoretical work on stochastic string excitation~\cite{bib1}, we present the design, fabrication, and spectral characterization of a custom-built noise-driven electroacoustic string instrument. The system implements open-loop stochastic electromagnetic actuation without feedback or pitch stabilization. We show that this excitation strategy produces a dense and uniformly distributed spectral regime that differs from conventional deterministic string excitation. This work contributes to a growing field of quantum music creation by offering a hybrid artistic–scientific platform with potential applications in live performance, experimental composition, and science education.}




\maketitle

\section{Introduction}\label{sec:introduction}

Electromagnetic and electroacoustic string instruments have been explored in the context of augmented musical interfaces, feedback systems, and hybrid acoustic--digital performance environments. In most existing approaches, electromagnetic actuation is employed either to reinforce deterministic string vibrations, to sustain tonal components, or to enable performative control through feedback and signal processing.

One representative example is the Magnetic Resonator Piano \cite{bib13}, which uses electromagnets to actuate piano strings with periodic waveforms, such as sinusoids, combined with phase-locked feedback control. The primary goal of this system is to sustain or extend traditional string excitation mechanisms, resulting in a stable tonal output characterized by a dominant fundamental frequency and well-defined harmonic overtones. Although the method expands the expressive capabilities of acoustic instruments, its excitation strategy remains fundamentally deterministic and pitch-oriented.

More recent instruments, such as the Sophtar \cite{bib12}, extend this paradigm through closed-loop feedback architectures, pressure-sensitive interfaces, embedded machine learning models, and advanced real-time DSP. In these systems, excitation signals are continuously derived from the instrument’s output and shaped by performer gestures or algorithmic processes.

Similarly, the Halldorophone \cite{bib11} operates through positive feedback, reinjecting amplified string signals into the instrument body to reinforce resonant modes, often emphasizing the fundamental frequency to generate sustained drones or controlled harmonic textures. While these instruments explore instability and nonlinear behavior, they remain feedback-driven systems in which the excitation is tightly coupled to the string’s own vibration.

In contrast to these approaches, the system presented in this work adopts an open-loop stochastic excitation paradigm, in which a white noise signal is applied directly to an electromagnetic actuator without feedback, phase synchronization, target frequency control, or embedded DSP. Rather than deriving or shaping the excitation signal from the instrument’s output, the resulting sonic behavior emerges from the physical interaction between broadband stochastic energy and the mechanical properties of the vibrating string. This excitation strategy produces a different vibrational regime, activating a broad and dense set of partials instead of reinforcing a number of dominant modes. As demonstrated by the spectral analysis presented later in this paper Figure~\ref{fig:spectral}, the resulting spectrum exhibits a flatter and more uniformly distributed energy profile, leading to a statistically structured and perceptually complex timbre.

\section{Objective and Conceptual Framework}\label{sec:framework}

This research aims to translate the theoretical analogy between the phase-space distributions of a classically driven harmonic oscillator under thermal fluctuations and those of a quantum harmonic oscillator in a coherent state into an audible phenomenon. The primary objective is to evaluate the timbral characteristics emerging from this type of excitation.

This experiment is conceptually grounded in recent developments in the theoretical and experimental study of harmonic oscillators in both classical and quantum regimes. In particular, this work draws on the insights developed in Classical and Quantum Parametric Phenomena~\cite{bib2} and in the study Stochastic Guitar~\cite{bib1}, both of which analyze how harmonic oscillators respond to parametric driving and stochastic perturbations. These studies reveal a striking mathematical similarity between the phase-space distributions of a classical oscillator under thermal fluctuations and that of a quantum harmonic oscillator in a coherent state.

Further experimental support for this analogy is found in Stochastic Guitar~\cite{bib1}, where the authors present measurements of guitar string oscillations under stochastic excitation. The stochastic force applied to the string is modeled as white noise of the form:

\begin{equation}
F(t) = \xi(t),
\end{equation}

\begin{equation}
\langle \xi(t) \rangle = 0,
\end{equation}

\begin{equation}
\langle \xi(t)\xi(t+\tau) \rangle = s^2 \delta(\tau),
\end{equation}

\vspace{0.5em}
\noindent
where $\langle \cdot \rangle$ denotes the long-time average, $\tau$ is the time delay, $\delta(\cdot)$ is the Dirac delta function, and $s^2$ represents the power spectral density of the noise force.

\vspace{0.5em}
\vspace{0.5em}
Rather than reproducing the full phase-space analysis presented in~\cite{bib1}, this work focuses on the physical implementation of stochastic excitation in a custom-built electroacoustic string instrument, and on the resulting acoustic and spectral behavior. The experimental realization and characterization of this system are presented in the following sections.

\section{Noise-Driven Instrument Design}\label{sec:instrument}

\begin{figure}[t]
\centering
\includegraphics[width=\linewidth]{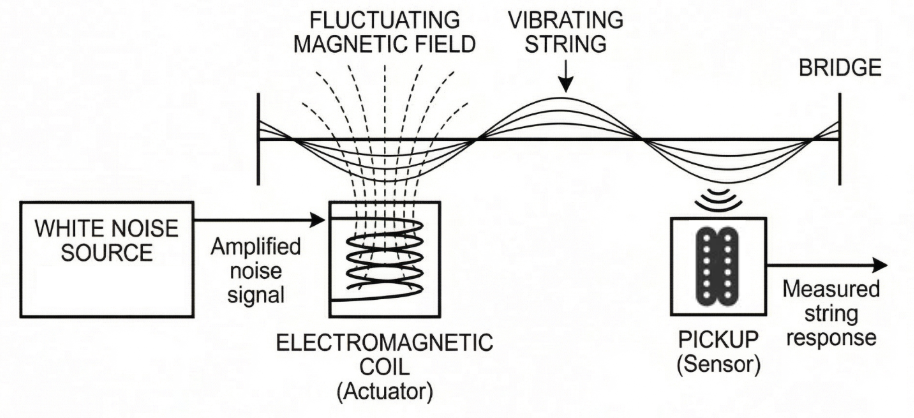}
\caption{Conceptual diagram of the open-loop noise-driven excitation and sensing architecture of the instrument.}
\label{fig:setup_conceptual}
\end{figure}

\subsection{Experimental Setup}\label{subsec:setup}

The experimental setup consists of a custom-built electroacoustic string instrument designed to enable controlled stochastic excitation and acoustic measurement. 

As illustrated in Figure~\ref{fig:setup_conceptual}, the system is based on the external injection of white noise into a string through electromagnetic actuation, allowing the resulting vibrational behavior to emerge from the physical interaction between the excitation field and the mechanical properties of the string.

A white noise signal is generated externally and routed through an audio interface and power amplifier to drive an electromagnetic actuator repurposed from a modified guitar pickup. The actuator induces string vibration via a fluctuating magnetic field, enabling contactless excitation of the instrument. The mechanical response of the string is captured by a second pickup positioned near the bridge, which converts the string motion into an electrical signal for subsequent spectral analysis.

A central challenge in the physical realization of the instrument is minimizing electromagnetic interference between the excitation and sensing elements. In particular, crosstalk—defined as the direct coupling of the excitation signal into the sensing pickup without mediation by string motion—must be reduced to ensure that the recorded signal accurately reflects the acoustic behavior of the string. To address this issue, two key design strategies were implemented: (i) the excitation coil was reengineered to improve magnetic field confinement and directionality, and (ii) the spatial layout of the instrument was optimized to increase the distance between the actuator and the sensing pickup.

\subsection{Electromagnetic Actuator: Coil Design and Construction}

The central element of the excitation system is a custom-designed electromagnetic actuator, repurposed from a conventional single-coil guitar pickup. To adapt the pickup for active excitation, the original winding was removed and the coil was rebuilt with parameters optimized for continuous electromagnetic actuation. This system inverts the function of a traditional pickup—from passive sensor to active exciter—allowing for programmable, contactless excitation.

The winding was constructed on the original bobbin, which features an internal rectangular geometry of approximately $6.0\,\mathrm{cm} \times 1.0\,\mathrm{cm}$. A copper wire of 28~AWG ($\approx 0.32\,\mathrm{mm}$ diameter) was selected for the electromagnetic actuator, as it offers a compromise between electrical resistance, thermal stability, and spatial efficiency. This gauge enables safe current levels for continuous operation while allowing a sufficient number of turns to achieve the desired magnetic field strength within the limited volume of the bobbin.

The resulting DC resistance of approximately $8\,\Omega$ ensures efficient power transfer from the amplifier without risk of overheating, making the actuator suitable for sustained electromagnetic excitation. The coil was therefore designed to present a target resistance of approximately $8\,\Omega$, matching the impedance required for optimal power transfer from the driving amplifier.

Using the resistivity of copper $\rho \approx 1.68 \times 10^{-8}\,\Omega\cdot\mathrm{m}$ and the cross-sectional area of 28~AWG wire, the required wire length was calculated to be approximately $38.3\,\mathrm{m}$. This corresponds to $\sim 255$ turns over the bobbin, assuming an average turn length of $15\,\mathrm{cm}$ adjusted for radial buildup.

To enhance the magnetic field strength and spatial focusing, the coil was magnetically coupled to three neodymium magnets, augmenting the original ceramic bar magnet of the pickup. This configuration improves the coupling efficiency between the actuator and the ferromagnetic guitar string, increasing vibrational amplitude and excitation stability.

\subsection{Instrument Fabrication and Structural Design}\label{subsec:fabrication}

\begin{figure}[htbp]
\centering
\includegraphics[width=\textwidth]{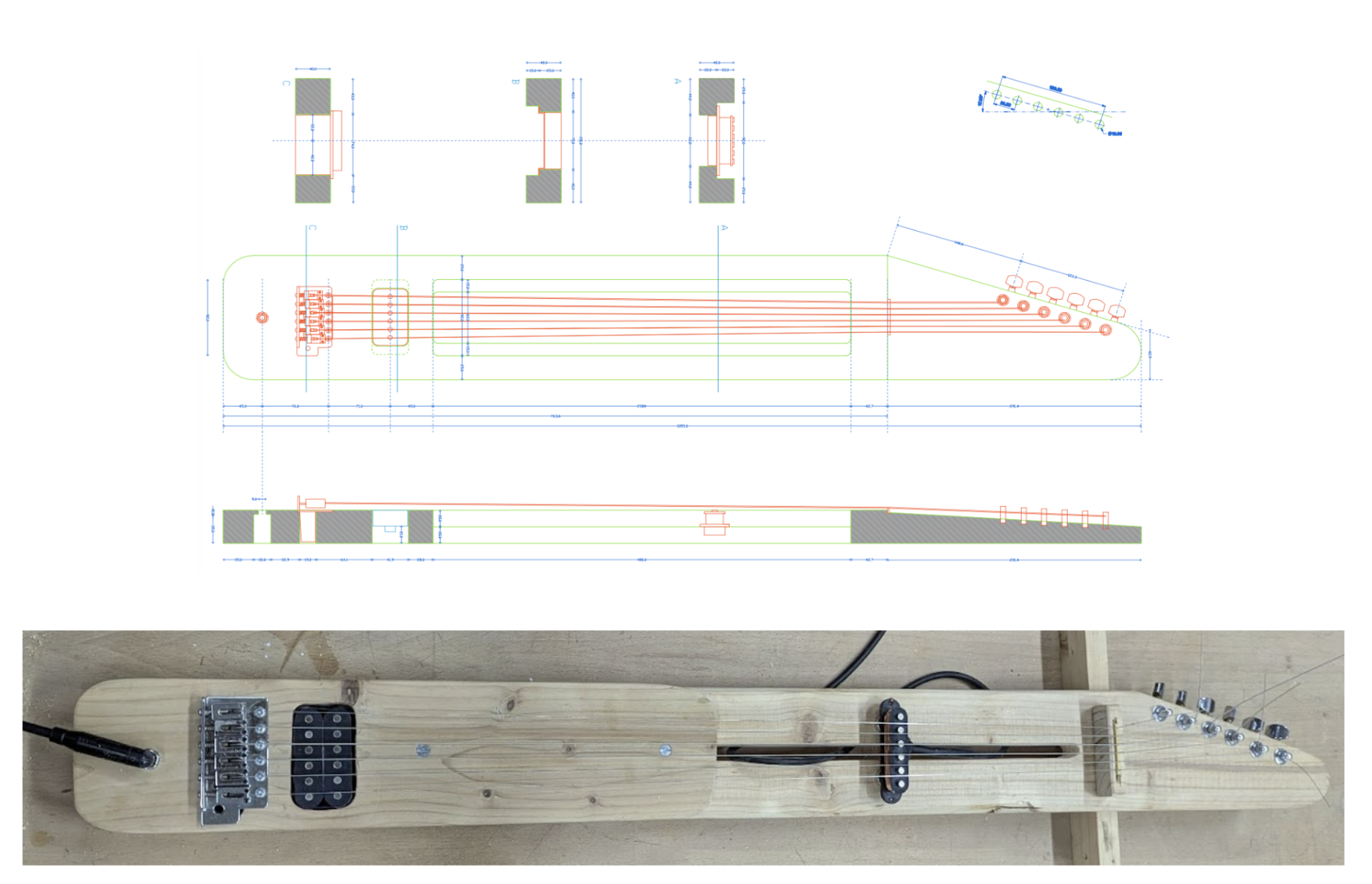}
\caption{Technical drawing and final prototype of the noise-driven instrument. 
Top: design schematic used for construction. 
Bottom: completed instrument integrating the electromagnetic actuator, pickup, and resonant string.}
\label{fig:instrument}
\end{figure}

Following the development of the electromagnetic actuation system, the physical structure of the instrument was designed using detailed technical schematics for the housing and mounting layout, as illustrated in Figure~\ref{fig:instrument}. The fabrication of the prototype was carried out in a workshop using a combination of cut wooden elements and manually assembled components. The design prioritized structural stability and access to the string and pickups. The resulting prototype maintains the tuning range of a conventional string instrument, while incorporating a programmable, noise-driven excitation mechanism optimized for experimental and performative applications.

\section{Spectral Analysis}\label{sec:spectral}

To evaluate the acoustic impact of stochastic excitation, a comparative spectral analysis was conducted between two activation modes of the instrument's string: (i) deterministic excitation using a conventional guitar pick, and (ii) noise-driven excitation via the electromagnetic actuator developed in this project. The goal was to characterize the resulting timbre by examining the distribution and density of spectral components in each case.

\begin{figure}[htbp]
\centering
\includegraphics[width=\textwidth]{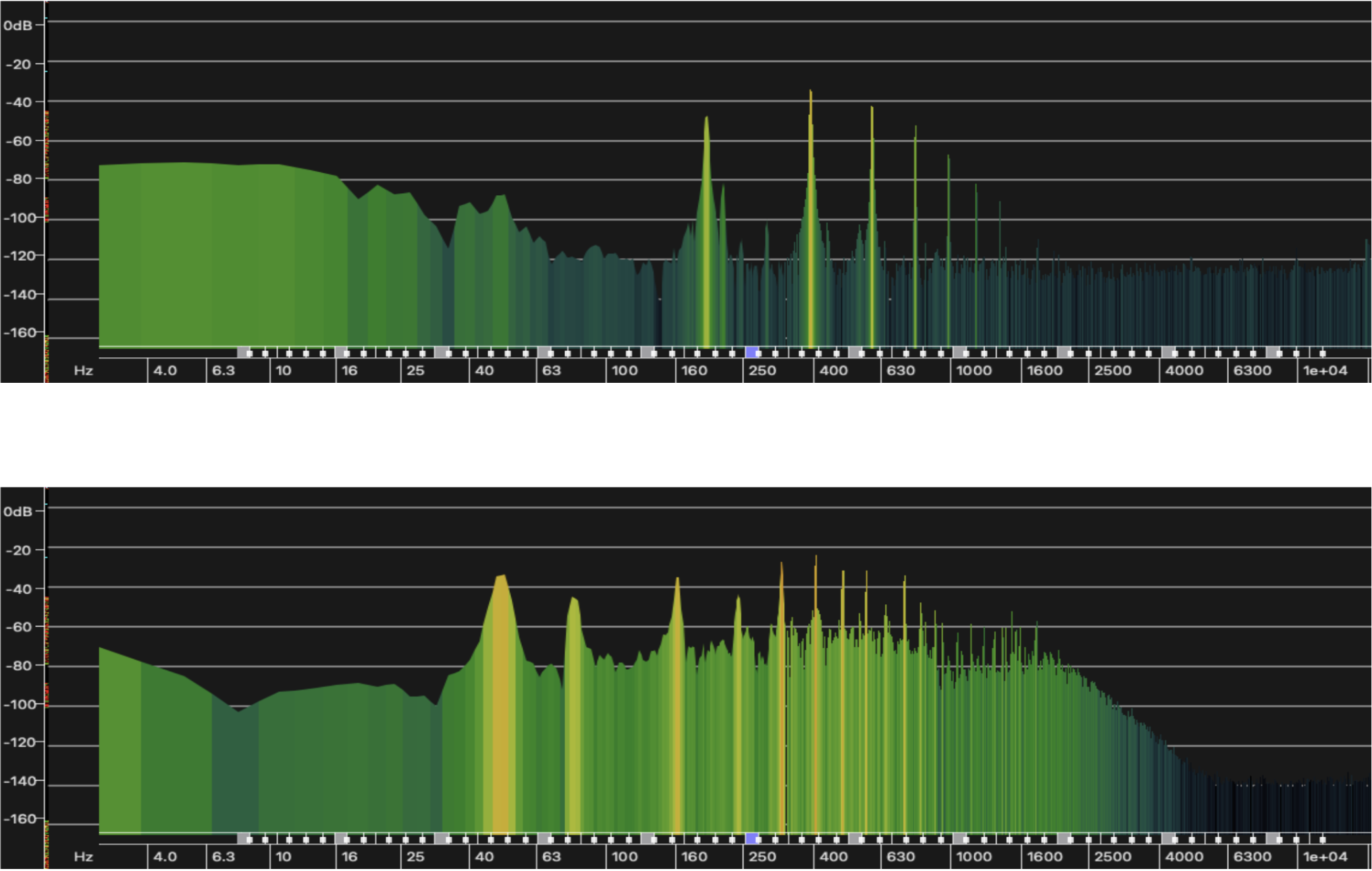}
\caption{Comparative frequency spectra of the vibrating string under two excitation methods. 
Top: excitation with a traditional guitar pick, showing a clear fundamental and harmonically structured overtones. 
Bottom: stochastic excitation with white noise, resulting in a denser and more uniformly distributed spectral profile, producing a richer and more complex timbre.}
\label{fig:spectral}
\end{figure}

As shown in Figure~\ref{fig:spectral}, the deterministic excitation produces a harmonic spectrum characterized by a dominant fundamental frequency and a series of regularly spaced overtones, consistent with classical models of string vibration. This spectral structure results in a stable and predictable timbre with clear pitch and low spectral complexity. In contrast, the spectrum resulting from stochastic excitation displays a much denser distribution of partials. Although the fundamental frequency remains visible, the surrounding harmonics are more numerous, closer in frequency, and more uniformly distributed across the spectrum. This leads to a richer and more complex timbre, with enhanced spectral energy in regions typically less active under conventional excitation.

\section{Discussion and Outlook}\label{sec:discussion}

This research demonstrates the potential of a multidisciplinary approach that bridges quantum physics and sound. By transforming quantum behaviors into sonic phenomena, the project establishes a novel framework for artistic exploration and materializes complex physical processes in audible form. The hybrid instrument—featuring a vibrating string excited by stochastic electromagnetic fields—introduces analog variability and performative richness, similar to traditional acoustic instruments. The physicality of the string and its resonance structure offers a contrast to purely digital systems, expanding the expressive palette and underscoring the relevance of material interaction in sound generation.
Future developments may include the implementation of distributed excitation systems using multi-coil arrays, allowing dynamic spatial control and spectral shaping, or the scaling of the architecture to excite an entire set of strings.
In contrast to existing electroacoustic string instruments based on feedback or deterministic excitation, the proposed instrument introduces a physically grounded stochastic excitation paradigm, offering a distinct spectral behavior and a new experimental platform for quantum-inspired sound research.
This project has demonstrated how quantum mechanical concepts can serve not only as metaphors but as operative frameworks for the design of new musical tools. The resulting system offers both a novel expressive medium for composers and performers, and a platform for exploring quantum phenomena through sound. Beyond its artistic contributions, this work lays the foundation for future developments that may deepen the integration between physics and musical practice.

From a broader perspective, this work is informed by mathematical structures that also appear in quantum mechanics, particularly the dynamics of the quantum harmonic oscillator. Recent studies have shown that mechanically implemented QHO-like systems can, under specific conditions, serve as building blocks for quantum information processing~\cite{bib3}. While the instrument presented here does not aim to implement qubits or computational functions, it operates within a classically driven systems whose stochastic phase-space behavior exhibits formal similarities to quantum coherent dynamics. In this sense, quantum models serve as a structural reference rather than a functional goal, suggesting future research directions in which such analogies may be further explored for the development of quantum-inspired musical instruments and sound practices.

 \bmhead{Acknowledgements}

We would like to thank Alexander Eichler (ETH Zurich) for his theoretical and practical support, as well as his group, especially Dr. Eric Clot, for guidance in the development of the electromagnetic excitation system. We also acknowledge Rober Martínez (Mutan Monkey Instruments, Barcelona) for technical assistance during prototyping, architect Armando Grand for the structural design of the instrument, and architect Jordi Quetglas for support during the fabrication phase.

This research was conducted at ICFO – Institut de Ciències Fotòniques as part of the BIST–UPF Master in Multidisciplinary Research in Experimental Sciences. The ICFO–Quantum Optics Theory group acknowledges support from the European Research Council (AdG NOQIA); MCIN/AEI (PGC2018-0910.13039/501100011033, CEX2019-000910-S/10.13039/501100011033, PID2019-106901GB-I00, PID2022-139099NB-I00, PRTR-C17.I1), QUANTERA DYNAMITE PCI2022-132919; QuantERA II Programme (Grant Agreement No. 101017733); the Ministry for Digital Transformation and of Civil Service of the Spanish Government through the Quantum Spain project; Fundació Cellex; Fundació Mir-Puig; Generalitat de Catalunya (CERCA programme and FEDER); Barcelona Supercomputing Center MareNostrum (FI-2023-3-0024); and the European Union’s Horizon Europe and Horizon 2020 programmes (HORIZON-CL4-2022-QUANTUM-02-SGA PASQuanS2.1, Grant No. 101113690; FET-OPEN OPTOlogic Grant No. 899794; QU-ATTO Grant No. 101168628; NeQST Grant No. 101080086). ICFO Internal “QuantumGaudi” project. Views and opinions expressed are those of the authors only and do not necessarily reflect those of the European Union, the European Commission, CINEA, or any other granting authority. Neither the European Union nor any granting authority can be held responsible for them. A.R.-P. acknowledges funding from ANR JCJC NEAR-2D (ANR-23-CE47-0015), Idex Welcome Package (UniCA), Fédération Doeblin FR2800 and UCA-JEDI (ANR-15-IDEX-01).

\bibliography{sn-bibliography}

\end{document}